# Proposal for an ultrasensitive spintronic strain and stress sensor

J. Atulasimha[1] and S. Bandyopadhyay[2,a]

Department of Mechanical Engineering

Virginia Commonwealth University

Richmond, VA 23284

Department of Electrical and Computer Engineering

Virginia Commonwealth University

Richmond, VA 23284

## ABSTRACT

We propose a spintronic strain/stress sensor capable of sensing strain with a sensitivity of ~ $10^{-13}/\sqrt{Hz}$ at room temperature with an active sensing area of ~ 1 cm$^2$ and power dissipation of ~ 1 watt. This device measures stress or strain by monitoring the change in the spin-polarized current in a parallel array of free standing nanowire spin valves when the array is subjected to compressive or tensile stress along the wires' length. Such a sensor can be fabricated using a variety of techniques involving nanolithography, self assembly and epitaxial growth.

[a] Corresponding author. E-mail: sbandy@vcu.edu

**I. Introduction**

Spintronic techniques are frequently used to sense magnetic fields with ultra-high sensitivity [1-5], but their application in sensing mechanical variables such as stress and strain is relatively unknown. In this paper, we show that a simple spintronic device, consisting of a parallel array of free-standing nanowire spin valves biased with a voltage source, can sense strain and stress by producing a change in the current flowing through it when strain or stress is applied. Strain can be sensed with a sensitivity of ~ $10^{-13}/\sqrt{Hz}$ at room temperature using an active area of ~ 1 cm$^2$ while dissipating only ~ 1 watt power. Reducing the sensing area by a factor of $N$ will degrade the sensitivity by a factor of $\sqrt{N}$ while reducing the power dissipation by a factor of $N$, so that a 100 μm × 100 μm active sensing area can still sense strain with $10^{-11}/\sqrt{Hz}$ sensitivity while consuming merely 100 μW. This sensitivity and power dissipation compare favorably with current technology for strain sensing [6, 7], which often use much larger devices that cannot be miniaturized. In the rest of this paper, we describe the theory underpinning this proposed spintronic sensor.

Consider a single free-standing nanowire spin-valve structure shown in Fig. 1 with current flowing along its length in the *x*-direction under an applied bias voltage *V*. The two ferromagnetic contacts of the spin valve are permanently magnetized by a magnetic field in the *x-z* plane, subtending an arbitrary angle $\theta$ with the z-axis. This field $\vec{H}_{applied}$ is applied by an external magnet and is constant. Let us now assume that one contact is nickel and the other is lanthanum strontium manganate (LSMO). In polycrystalline form, nickel is magnetostrictive with negative magnetostriction coefficient [8], while LSMO is magnetostrictive with positive magnetostriction coefficient [9]. The contacts are assumed to be small enough to act as single-domain nanomagnets but not so small as to be super-paramagnetic at room temperature. They have square shape in the *x-z* plane and their thickness (in the *y*-direction) is assumed to be much smaller than the linear dimension in the *x*- or *z*-direction. These requirements are fulfilled if the dimensions are 30 nm × 10 nm × 30 nm. Note that since nickel and LSMO have opposite signs of spin polarization at the Fermi level, the same magnetic field places them in the "anti-parallel" configuration whereby the majority spin in one magnet is the minority spin in the other and vice versa. As a result, if everything were ideal (e.g. 100% efficient spin injection and detection and no spin relaxation), then no current will flow through the wire since the spins injected by one contact will be completely blocked by the other. It is as if the wire has infinite resistance.

If stress is applied on the wire in the *x*-direction, it will cause the magnetizations of both contacts to rotate by small angles $\Delta\theta_1$ and $\Delta\theta_2$ over the *x-z* plane, but in *opposite* directions, since the signs of the magnetostriction coefficients of the two materials are opposite. This is shown in Fig. 1. Any *y*-component of rotation is not favored since the thickness of the contacts (in the *y*-direction) is much smaller than the linear dimensions in the *x-z* plane. Therefore, the effect of stress is to make the two contacts deviate from being exactly "anti-parallel" by an angle $|\Delta\theta_1|+|\Delta\theta_2|$. This will cause some current to flow through the wire. Even if the contacts are non-ideal and do not inject and detect spin with 100% efficiency, stress will cause a *change* in the current flowing through the wire, which can be measured with electrometers or a Wheatstone bridge network, and will yield a measurement of the applied stress as well as the resulting strain in any of the layers. This is the principle of spintronic stress or strain sensing.

## II. Theory

In this section, we will show: (1) that stress applied along the length of the wire (*x*-direction) will rotate both contacts' magnetizations from their initial orientation, and (2) that this rotation will cause a measurable change in the spin-polarized current flowing in the wire.

To establish that the magnetization vector of either magnetostrictive contact will rotate under stress, assume that the contact is a single-domain ferromagnet, and subjected to a tensile or compressive stress $\sigma$ along the *x*-direction. In our convention, tensile stress is positive and compressive stress is negative. The magnetic energy of such a stressed nanomagnet, as a function of the angle $\Delta\theta$ that it subtends with the applied magnetic field $H_{applied}$, can be written as [10]:

$$E(\Delta\theta) = \left[-\mu_0 H_{applied} M_s \cos(\Delta\theta) - \frac{3}{2}\lambda_s \sigma \cos^2\varphi\right]v$$
$$= -\mu_0 H_{applied} M_s \cos(\Delta\theta) v - \frac{3}{2}\lambda_s \sigma \cos^2\left(\frac{\pi}{2} - \theta - \Delta\theta\right)v \quad , \quad (1)$$

where *v* is the volume of the nanomagnet, $\varphi = \pi/2 - \theta - \Delta\theta$ is the angle between the magnetization vector and the stress, $\mu_0$ is the permeability of free space, $(3/2)\lambda_s$ is the saturation magnetostriction of the contacts, and $M_s$ is the saturation magnetization moment. Here, we have included the magnetization energy (first term) and the stress anisotropy energy (second term), while neglecting the shape anisotropy energy. The shape anisotropy along the *x-z* plane is very small owing to the small thickness in the *y*-direction and the fact that the shape of magnet in the *x-z* plane is a square. Therefore, any shape anisotropy energy will be independent of $\Delta\theta$ and hence can be neglected. We have also ignored magnetocrystalline anisotropy since our contacts will be polycrystalline.

Note from Equation (1) that

$$E(\Delta\theta) - E(0) = \mu_0 H_{applied} M_s \left[1 - \cos(\Delta\theta)\right] v + \frac{3}{2}\lambda_s \sigma \left[\sin^2\theta - \sin^2(\theta + \Delta\theta)\right] v. \quad (2)$$

Clearly, as long as the right hand side of the above equation is negative, the contact can lower its energy by rotating its magnetization in the *x-z* plane by an angle $\Delta\theta$ away from the initial orientation. Therefore, in order for rotation to occur, we need to satisfy the inequality

$$\frac{3}{2}\lambda_s \sigma \left[\sin^2(\theta + \Delta\theta) - \sin^2\theta\right] > \mu_0 H_{applied} M_s \left[1 - \cos(\Delta\theta)\right]. \quad (3)$$

Since the right hand side is always positive, a negative $\lambda_s \sigma$ product (negative magnetostriction and tensile stress, or positive magnetostriction and compressive stress) will result in negative $\Delta\theta$ (rotation towards the z-axis away from the x-axis), while a positive $\lambda_s \sigma$ product will result in positive $\Delta\theta$ (rotation away from the z-axis towards the x-axis). As a result, the magnetizations of nickel and LSMO contacts – which have opposite signs of $\lambda_s$ – will rotate in opposite directions when subjected to the same stress.

Because $\Delta\theta$ will be small, we can recast Equation (3) as

$$\frac{3}{2}\lambda_s \sigma \left[\sin(2\theta)\Delta\theta + \cos^2\theta (\Delta\theta)^2\right] > \mu_0 H_{applied} M_s \frac{1}{2}(\Delta\theta)^2. \quad (4)$$

Equation (4) yields some interesting insights on the effect of stress in rotating the magnetization and the importance of the angle θ between the stress and field directions. Consider, the case when stress and the magnetic field are collinear, i.e. θ=0. Equation (4) can now be simplified to:

$$\frac{3}{2}\lambda_s \sigma [\Delta\theta]^2 > \mu_0 H_{applied} M_s [\Delta\theta]^2 \tag{4a}$$

This inequality shows that no rotation will occur until $\frac{3}{2}\lambda_s \sigma > \mu_0 H_{applied} M_s$, after which there will be a sudden rotation through a large angle. Clearly, there can be no rotation until the stress anisotropy energy has completely overcome the magnetic energy. However, the scenario is completely different if stress is applied at 45º to the magnetic field. For θ=45º, Equation (4) simplifies to:

$$\frac{3}{2}\lambda_s \sigma \left[\Delta\theta + \frac{1}{2}(\Delta\theta)^2\right] > \mu_0 H_{applied} M_s [\Delta\theta]^2 \tag{4b}$$

For small angles $\Delta\theta \gg (\Delta\theta)^2$ so we only need $\frac{3}{2}\lambda_s \sigma [\Delta\theta] > 0$ to be satisfied. This implies that a negative $\lambda_s \sigma$ product (negative magnetostriction and tensile stress, or positive magnetostriction and compressive stress) will definitely result in negative $\Delta\theta$ (rotation towards the z-axis away from the x-axis), while a positive $\lambda_s \sigma$ product will result in positive $\Delta\theta$ (rotation away from the z-axis towards the x-axis).

Once Equation (3) or (4) is satisfied and rotation takes place, the next step is to find $\Delta\theta$ for a given stress σ. For this purpose, we have to minimize the energy given by Equation (1) or Equation (2). We do this by setting $\partial E(\Delta\theta)/\partial(\Delta\theta) = 0$, or $\partial[E(\Delta\theta) - E(0)]/\partial(\Delta\theta) = 0$, which yields the condition

$$\mu_0 H_{applied} M_s \sin(\Delta\theta) - \frac{3}{2}\lambda_s \sigma \sin(2\theta + 2\Delta\theta)$$
$$= \mu_0 H_{applied} M_s \sin(\Delta\theta) - \frac{3}{2}\lambda_s \sigma [\sin(2\theta)\cos(2\Delta\theta) + \cos(2\theta)\sin(2\Delta\theta)]. \quad (5)$$
$$= 0$$

Once again, since $\Delta\theta$ is small, we set $\lim_{\Delta\theta \to 0} \sin(2\Delta\theta) \approx 2\Delta\theta$ and $\lim_{\Delta\theta \to 0} \cos(2\Delta\theta) \approx 1$ to get:

$$\mu_0 H_{applied} M_s \Delta\theta \approx \frac{3}{2}\lambda_s \sigma [\sin(2\theta) + 2\cos(2\theta)\Delta\theta], \quad (6)$$

which yields

$$\Delta\theta \approx \frac{(3/2)\lambda_s \sigma \sin(2\theta)}{\mu_0 H_{applied} M_s - 3\lambda_s \sigma \cos(2\theta)}. \quad (7)$$

To find the optimum value of $\theta$ (orientation of the applied magnetic field in the x-z plane) which will yield the largest $\Delta\theta$, we set $\frac{\partial \Delta\theta}{\partial \theta} = 0$ and solve for $\theta$, which yields the optimum value as

$$\theta_{opt} = \frac{1}{2}\cos^{-1}\left[\frac{3\lambda_s \sigma}{\mu_0 H_{applied} M_s}\right] \quad (8)$$

Unfortunately, the optimum value of $\theta$ depends on the stress $\sigma$ which is a variable quantity. Since we cannot continually reorient the applied magnetic field depending on the stress encountered, we cannot operate at optimum $\theta$. Furthermore, we would like to make the stress-induced rotation linearly proportional to stress. This is achieved if we simply choose $\theta = \pi/4$, which then yields (from Equation (7))

$$\Delta\theta = \frac{(3/2)\lambda_s \sigma}{\mu_0 H_{applied} M_s} \qquad (9)$$

It is easy to verify that the second derivative $\partial^2 E(\Delta\theta)/\partial(\Delta\theta)^2$ or $\partial^2 [E(\Delta\theta)-E(0)]/\partial(\Delta\theta)^2$ evaluated at the value of $\Delta\theta$ given by Equation (9) is $\mu_0 H_{applied} M_s + 4[(3/2)\lambda_s \sigma]^2 / \mu_0 H_{applied} M_s$, which is always positive, so that Equation (9) indeed yields the value of $\Delta\theta$ corresponding to the minimum energy configuration (ground state) if we had oriented the applied magnetic field at $45^0$ with the *z*-axis.

Substitution of Equation (9) in Equation (4) with $\theta = \pi/4$ shows that Equation (4) is always satisfied by the value of $\Delta\theta$ given by Equation (9). Therefore, a contact's magnetization invariably rotates by an angle $\Delta\theta$ when stressed.

Next, we need to establish that the rotation of the nickel contact's magnetization will induce a change in the spin-polarized current flowing in the wire under an applied electrical bias. To keep the algebra simple, we will assume that only the lowest electronic subband in the semiconductor quantum wire is occupied by electrons.

We will work out the formulation for arbitrary values of $\theta$ and not just for $\theta = 45^0$. The purpose of this is to show that the change in the current caused by stress depends only on $\Delta\theta$ and is independent of $\theta$. Therefore, even if the magnetic field is slightly misaligned from the $45^0$ orientation, the change in the current caused by stress will not be affected. In other words, the sensor is fault-tolerant.

Since nickel has a negative spin polarization at the Fermi level, its majority spin's orientation points anti-parallel to the magnetization or the applied magnetic field when there is no stress. On the other hand, LSMO has positive spin polarization at the Fermi level. Hence, its majority spin will point parallel to the direction of the magnetic field. Therefore, when stress is applied and a contact's magnetization rotates by an angle $\Delta\theta$, the eigenspinors within the two contacts at the Fermi energy can be written in the Bloch sphere representation as [11]

$$[\Phi]_{maj}^{Nickel} = \begin{bmatrix} \sin\left(\frac{\theta+\Delta\theta_1}{2}\right) \\ -e^{i\phi}\cos\left(\frac{\theta+\Delta\theta_1}{2}\right) \end{bmatrix}; \quad [\Phi]_{min}^{Nickel} = \begin{bmatrix} \cos\left(\frac{\theta+\Delta\theta_1}{2}\right) \\ e^{i\phi}\sin\left(\frac{\theta+\Delta\theta_1}{2}\right) \end{bmatrix}$$

$$[\Phi]_{maj}^{LSMO} = \begin{bmatrix} \cos\left(\frac{\theta+\Delta\theta_2}{2}\right) \\ e^{i\phi}\sin\left(\frac{\theta+\Delta\theta_2}{2}\right) \end{bmatrix}; \quad [\Phi]_{min}^{LSMO} = \begin{bmatrix} \sin\left(\frac{\theta+\Delta\theta_2}{2}\right) \\ -e^{i\phi}\cos\left(\frac{\theta+\Delta\theta_2}{2}\right) \end{bmatrix}, \quad (10)$$

where the subscripts 'maj' and 'min' indicate majority and minority spins in the contacts and $\phi$ is the azimuthal angle of the magnetization vector, which in our case, is zero. Here, $\Delta\theta_1$ and $\Delta\theta_2$ are the rotations of the nickel and LSMO contacts, respectively. They will be of opposite signs.

The magnetic field $H_{applied}$ will result in a small spin splitting in the wire due to the Zeeman interaction Hamiltonian $-(g/2)\mu_B\mu_0\vec{H}_{applied}\bullet\vec{\sigma}$, where $\mu_B$ is the Bohr magneton, $g$ is the Lande g-factor in the wire and $\vec{\sigma}$ is the Pauli spin matrix. The dispersion relations of the resulting spin-split levels will be

$$E_\pm = \frac{\hbar^2 k^2}{2m^*} \pm \frac{g}{2}\mu_B\mu_0 H_{applied}, \quad (11)$$

while the eigenspinors in these levels will be (assuming a positive g-factor so that the majority spins in the wire line up parallel to the applied magnetic field)

$$[\Psi]_- = \begin{bmatrix} \cos\left(\frac{\theta}{2}\right) \\ e^{i\phi}\sin\left(\frac{\theta}{2}\right) \end{bmatrix}; \quad [\Psi]_+ = \begin{bmatrix} \sin\left(\frac{\theta}{2}\right) \\ -e^{i\phi}\cos\left(\frac{\theta}{2}\right) \end{bmatrix}. \quad (12)$$

Here, we have neglected spin-orbit interaction which can cause additional spin splitting. In direct-gap semiconductors, whose conduction band minimum is in the $\Gamma$-valley, there is no intrinsic spin-orbit interaction because the Bloch states are $|S\rangle$-type. There can however be both Rashba and Dresselhaus interactions arising from structural and bulk inversion asymmetries, respectively. We ignore Rashba interaction since that requires a symmetry breaking electric field, which we do not have, and we ignore Dresselhaus interaction since it is weak in most semiconductors.

For purposes of the following analysis, we will assume that the polarity of the battery is such that the injecting contact is nickel and the detecting contact is LSMO (see Fig. 1). Of course nothing will change if we had assumed the opposite. When an electron is injected into the wire from the majority spin band in the nickel contact with spin state $[\Phi]_{maj}^{Nickel}$, it will couple into the eigenspinors in the wire with coupling coefficients $C_1$ and $C_2$ according to

$$[\Phi]_{maj}^{Nickel} = \begin{bmatrix} \sin\left(\frac{\theta+\Delta\theta_1}{2}\right) \\ -e^{i\phi}\cos\left(\frac{\theta+\Delta\theta_1}{2}\right) \end{bmatrix}$$

$$= C_1[\Psi]_- + C_2[\Psi]_+ = C_1 \begin{bmatrix} \cos\left(\frac{\theta}{2}\right) \\ e^{i\phi}\sin\left(\frac{\theta}{2}\right) \end{bmatrix} + C_2 \begin{bmatrix} \sin\left(\frac{\theta}{2}\right) \\ -e^{i\phi}\cos\left(\frac{\theta}{2}\right) \end{bmatrix}, \quad (13)$$

which immediately yields

$$\begin{aligned} C_1 &= \sin\left(\frac{\Delta\theta_1}{2}\right) \\ C_2 &= \cos\left(\frac{\Delta\theta_1}{2}\right) \end{aligned}. \quad (14)$$

If the injected electron reaches the detecting contact (LSMO) without suffering spin relaxation in the wire, then its eigenspinor at the right contact will be

$$\begin{aligned}
[\Psi]_{majority}^{LSMO} &= C_1 \begin{bmatrix} \cos\left(\frac{\theta}{2}\right) \\ e^{i\phi}\sin\left(\frac{\theta}{2}\right) \end{bmatrix} e^{ik_1 L} + C_2 \begin{bmatrix} \sin\left(\frac{\theta}{2}\right) \\ -e^{i\phi}\cos\left(\frac{\theta}{2}\right) \end{bmatrix} e^{ik_2 L} \\
&= \sin\left(\frac{\Delta\theta_1}{2}\right) \begin{bmatrix} \cos\left(\frac{\theta}{2}\right) \\ e^{i\phi}\sin\left(\frac{\theta}{2}\right) \end{bmatrix} e^{ik_1 L} + \cos\left(\frac{\Delta\theta_1}{2}\right) \begin{bmatrix} \sin\left(\frac{\theta}{2}\right) \\ -e^{i\phi}\cos\left(\frac{\theta}{2}\right) \end{bmatrix} e^{ik_2 L}
\end{aligned}, \quad (15)$$

where $k_1$ and $k_2$ are the wavevectors in the two spin split bands at the Fermi energy $E_F$. These two quantities are found from the dispersion relations shown in Fig. 2:

$$E_F = \frac{\hbar^2 k_1^2}{2m^*} - \frac{g}{2}\mu_B\mu_0 H_{applied} = \frac{\hbar^2 k_2^2}{2m^*} + \frac{g}{2}\mu_B\mu_0 H_{applied}. \quad (16)$$

Similarly, an electron injected from the minority spin band in the injecting contact with spin state $[\Phi]_{\text{min}}^{\text{Nickel}}$ will couple into the eigenspinors in the wire with coupling coefficients $C_1'$ and $C_2'$ given by

$$[\Phi]_{\text{min}}^{\text{Nickel}} = \begin{bmatrix} \cos\left(\dfrac{\theta + \Delta\theta_1}{2}\right) \\ e^{i\phi} \sin\left(\dfrac{\theta + \Delta\theta_1}{2}\right) \end{bmatrix}$$

$$= C_1'[\Psi]_- + C_2'[\Psi]_+ = C_1' \begin{bmatrix} \cos\left(\dfrac{\theta}{2}\right) \\ e^{i\phi} \sin\left(\dfrac{\theta}{2}\right) \end{bmatrix} + C_2' \begin{bmatrix} \sin\left(\dfrac{\theta}{2}\right) \\ -e^{i\phi} \cos\left(\dfrac{\theta}{2}\right) \end{bmatrix}, \quad (17)$$

which yields

$$\begin{aligned} C_1' &= \cos\left(\dfrac{\Delta\theta_1}{2}\right) \\ C_2' &= -\sin\left(\dfrac{\Delta\theta_1}{2}\right) \end{aligned}. \quad (18)$$

If this injected electron reaches the detecting contact (LSMO) without suffering spin relaxation, then its eigenspinor at the right contact will be

$$[\Psi]_{\text{minority}}^{\text{LSMO}} = C_1' \begin{bmatrix} \cos\left(\dfrac{\theta}{2}\right) \\ e^{i\phi} \sin\left(\dfrac{\theta}{2}\right) \end{bmatrix} e^{ik_1 L} + C_2' \begin{bmatrix} \sin\left(\dfrac{\theta}{2}\right) \\ -e^{i\phi} \cos\left(\dfrac{\theta}{2}\right) \end{bmatrix} e^{ik_2 L}$$

$$= \cos\left(\dfrac{\Delta\theta_1}{2}\right) \begin{bmatrix} \cos\left(\dfrac{\theta}{2}\right) \\ e^{i\phi} \sin\left(\dfrac{\theta}{2}\right) \end{bmatrix} e^{ik_1 L} - \sin\left(\dfrac{\Delta\theta_1}{2}\right) \begin{bmatrix} \sin\left(\dfrac{\theta}{2}\right) \\ -e^{i\phi} \cos\left(\dfrac{\theta}{2}\right) \end{bmatrix} e^{ik_2 L} \quad (19)$$

In the preceding analysis, we assumed that the injected spin does not relax within the semiconductor wire. In most semiconductors, spin relaxation is caused primarily by the D'yakonov-Perel' [12] and the Elliott-Yafet [13] modes. The former is absent in a quantum wire with a single occupied subband [14, 15] and the latter is also very weak in most semiconductors [16]. Therefore, it is permissible to ignore spin relaxation as long as the wire is not too long (shorter than a few μm). Note that transport in the semiconductor layer does not have to be ballistic; the electron can suffer frequent momentum randomizing collisions with phonons and impurities, but they do not matter unless they relax spin. This is a consequence of the strictly one-dimensional nature of the wire.

Now, consider the case when a majority spin injected from the injecting contact (nickel) tries to transmit into the majority spin band of the detecting contact (LSMO). The corresponding transmission amplitude is

$$T_{\text{maj-maj}} = \left\{[\Phi]_{\text{maj}}^{\text{LSMO}}\right\}^{\dagger} [\Psi]_{\text{majority}}^{\text{LSMO}} = e^{ik_1 L} \sin\left(\frac{\Delta\theta_1}{2}\right)\cos\left(\frac{\Delta\theta_2}{2}\right) - e^{ik_2 L} \sin\left(\frac{\Delta\theta_2}{2}\right)\cos\left(\frac{\Delta\theta_1}{2}\right), \quad (20)$$

where the "dagger" represents hermitian adjoint. The associated transmission probability is

$$\left|T_{\text{maj-maj}}\right|^2 = \sin^2\left(\frac{\Delta\theta_1 - \Delta\theta_2}{2}\right) + \sin(\Delta\theta_1)\sin(\Delta\theta_2)\sin^2\frac{\alpha}{2}, \text{ where } \alpha = (k_2 - k_1)L.$$

Similarly, if a majority spin is injected from the injecting contact and tries to transmit into the minority spin band in the detecting contact, then the corresponding transmission amplitude is

$$T_{\text{maj-min}} = \left\{[\Phi]_{\text{min}}^{\text{LSMO}}\right\}^{\dagger} [\Psi]_{\text{majority}}^{\text{LSMO}} = e^{ik_1 L} \sin\left(\frac{\Delta\theta_1}{2}\right)\sin\left(\frac{\Delta\theta_2}{2}\right) + e^{ik_2 L} \cos\left(\frac{\Delta\theta_2}{2}\right)\cos\left(\frac{\Delta\theta_1}{2}\right) \quad (21)$$

and the associated transmission probability is

$$|T_{\text{maj-min}}|^2 = \cos^2\left(\frac{\Delta\theta_1 - \Delta\theta_2}{2}\right) - \sin(\Delta\theta_1)\sin(\Delta\theta_2)\sin^2\frac{\alpha}{2}.$$

Next, we will consider cases when a minority spin is injected from the injecting contact. If it tries to transmit into the majority spin band in the detecting contact, the transmission amplitude will be

$$T_{\text{min-maj}} = \left\{[\Phi]_{\text{maj}}^{\text{LSMO}}\right\}^\dagger [\Psi]_{\text{minority}}^{\text{LSMO}} = e^{ik_2 L}\sin\left(\frac{\Delta\theta_1}{2}\right)\sin\left(\frac{\Delta\theta_2}{2}\right) + e^{ik_1 L}\cos\left(\frac{\Delta\theta_2}{2}\right)\cos\left(\frac{\Delta\theta_1}{2}\right) \quad (22)$$

and the corresponding transmission probability is

$$|T_{\text{min-maj}}|^2 = \cos^2\left(\frac{\Delta\theta_1 - \Delta\theta_2}{2}\right) - \sin(\Delta\theta_1)\sin(\Delta\theta_2)\sin^2\frac{\alpha}{2}.$$

On the other hand, if a minority spin injected from the injecting contact tries to transmit into the minority spin band in the detecting contact, the corresponding transmission amplitude will be

$$T_{\text{min-min}} = \left\{[\Phi]_{\text{min}}^{\text{LSMO}}\right\}^\dagger [\Psi]_{\text{minority}}^{\text{LSMO}} = e^{ik_1 L}\sin\left(\frac{\Delta\theta_2}{2}\right)\cos\left(\frac{\Delta\theta_2}{2}\right) - e^{ik_2 L}\sin\left(\frac{\Delta\theta_1}{2}\right)\cos\left(\frac{\Delta\theta_2}{2}\right) \quad (23)$$

and the associated transmission probability will be

$$|T_{\text{min-min}}|^2 = \sin^2\left(\frac{\Delta\theta_1 - \Delta\theta_2}{2}\right) + \sin(\Delta\theta_1)\sin(\Delta\theta_2)\sin^2\frac{\alpha}{2}.$$

Let us now assume that the spin injection efficiency at the injecting contact is $\varepsilon_l$ and the spin detection efficiency at the detecting contact is $\varepsilon_r$. Then the various injection/detection scenarios take place with relative probabilities given by

$$\text{maj-min} \to \frac{(1+\varepsilon_l)(1-\varepsilon_r)}{4}; \quad \text{maj-maj} \to \frac{(1+\varepsilon_l)(1+\varepsilon_r)}{4}$$
$$\text{min-maj} \to \frac{(1-\varepsilon_l)(1+\varepsilon_r)}{4}; \quad \text{min-min} \to \frac{(1-\varepsilon_l)(1-\varepsilon_r)}{4}. \tag{24}$$

Since the majority and minority spin states in the contacts are mutually orthogonal, the weighted transmission probability, considering all possible scenarios, is

$$|T|^2 = \frac{(1+\varepsilon_l)(1+\varepsilon_r)}{4}|T_{\text{maj-maj}}|^2 + \frac{(1+\varepsilon_l)(1-\varepsilon_r)}{4}|T_{\text{maj-min}}|^2 + \frac{(1-\varepsilon_l)(1+\varepsilon_r)}{4}|T_{\text{min-maj}}|^2 + \frac{(1-\varepsilon_l)(1-\varepsilon_r)}{4}|T_{\text{min-min}}|^2 \tag{25}$$
$$= \frac{1}{2}\left[1 - \varepsilon_l\varepsilon_r \cos(\Delta\theta_1 - \Delta\theta_2) + 2\varepsilon_l\varepsilon_r \sin(\Delta\theta_1)\sin(\Delta\theta_2)\sin^2\frac{\alpha}{2}\right].$$

Note that the total transmission probability is *not* energy independent because of the last term that involves $\alpha$, an energy-dependent quantity.

In the linear response regime, the spin-polarized current in the quantum wire is related to the applied bias between the contacts according to [17]

$$I = \frac{e^2}{h}V\int_0^\infty |T(E)|^2 dE\left(-\frac{\partial f(E)}{\partial E}\right)$$
$$= \frac{e^2}{4h}\left[1 - \varepsilon_l\varepsilon_r \cos(\Delta\theta_1 - \Delta\theta_2)\right]\left[1 + \tanh\left(\frac{E_F}{2kT}\right)\right]V + \frac{e^2}{h}V\varepsilon_l\varepsilon_r \sin(\Delta\theta_1)\sin(\Delta\theta_2)\int_0^\infty \sin^2\frac{\alpha}{2} dE\left(-\frac{\partial f(E)}{\partial E}\right), \tag{26}$$
$$= \frac{e^2}{4h}\left[1 - \varepsilon_l\varepsilon_r \cos(\Delta\theta_1 - \Delta\theta_2)\right]\left[1 + \tanh\left(\frac{E_F}{2kT}\right)\right]V + \frac{e^2}{h}V\Gamma\varepsilon_l\varepsilon_r \sin(\Delta\theta_1)\sin(\Delta\theta_2)$$

where $V$ is the voltage applied between the contacts, $f(E)$ is the electron occupation probability in the contacts ($E$ is the electron energy) and $\Gamma = \int_0^\infty dE \sin^2\frac{\alpha(E)}{2}\left(-\frac{\partial f(E)}{\partial E}\right)$. Since the contacts are in local thermodynamic equilibrium, we assumed $f(E)$ to be Fermi-Dirac and $E_F$ is the Fermi energy in the contacts.

When stress is applied to rotate the magnetizations of the contacts, the current through the sensor changes by an amount $\Delta I = I(\Delta\theta_1 - \Delta\theta_2) - I(0)$, which yields

$$\Delta I = \frac{e^2}{4h}\varepsilon_l\varepsilon_r \left\{\left[1-\cos(\Delta\theta_1 - \Delta\theta_2)\right]\left[1+\tanh\left(\frac{E_F}{2kT}\right)\right] - 4\Gamma\sin(\Delta\theta_1)\sin(\Delta\theta_2)\right\}V$$
$$\approx \frac{e^2}{2h}\varepsilon_l\varepsilon_r \sin^2\left(\frac{\Delta\theta_1 - \Delta\theta_2}{2}\right)\left[1+\tanh\left(\frac{E_F}{2kT}\right)\right]V \quad (27)$$

since $\Gamma$ is typically small. We will justify this later.

Note that $\Delta I$ is independent of $\theta$ which is the orientation of the magnetic field $H_{applied}$ on the $x$-$z$ plane. Hence a slight misalignment from the $45^0$ orientation will not affect $\Delta I$. As stated before, this makes the sensor *fault-tolerant*. Note also that if $\varepsilon_l = \varepsilon_r = 1$, i.e. if we have perfect spin injection and detection, then $I(0) = 0$, which is expected since the two magnetic contacts are in the anti-parallel configuration. In this case, there will be no standby current (when no stress is applied) and hence no standby power dissipation.

### III. Calculation of the sensitivity

We will now calculate the smallest $\Delta\theta = \Delta\theta_1 - \Delta\theta_2$ that we can detect by measuring the corresponding change in the spin-polarized current $\Delta I$. This will yield the sensitivity of the sensor.

Note that $\Delta I$ is the sensor's signal current $I_s$. We need to find the noise current $I_n = \sqrt{\langle\Delta I_s^2\rangle}$. At room temperature, the major source of noise in a semiconductor wire is usually Johnson noise.

The Johnson noise current in a single wire is $I_n = \sqrt{4kTG\Delta f}$, where $G$ is the conductance of the wire when the current through it $I_s$. In other words, $G = I_s/V$. In a quantum wire, Johnson noise may be suppressed because of phonon confinement [18], but we will be conservative and ignore such suppression.

The signal-to-noise ratio of a single wire will then be

$$(SNR)_1 = \frac{I_s}{I_n} = \frac{I_s}{\sqrt{4kT\Delta fG}} = \frac{(e^2/2h)\varepsilon_l\varepsilon_r \sin^2\left(\frac{\Delta\theta}{2}\right)\left[1+\tanh\left(\frac{E_F}{2kT}\right)\right]V}{\sqrt{4kT\Delta f(e^2/2h)\varepsilon_l\varepsilon_r \sin^2\left(\frac{\Delta\theta}{2}\right)\left[1+\tanh\left(\frac{E_F}{2kT}\right)\right]}}$$

$$= \frac{(e^2/h)^{1/2}\sqrt{\varepsilon_l\varepsilon_r}\sin\left(\frac{\Delta\theta}{2}\right)\sqrt{1+\tanh\left(\frac{E_F}{2kT}\right)}V}{2\sqrt{2kT\Delta f}} \quad (28)$$

The G in the denominator of equation (28) was approximately calculated as $G = \Delta I/V$ which is accurate only when I(0) = 0. If spin injection and detection efficiency $\varepsilon_r$ and $\varepsilon_l$ are not close to one then $G = [I(0)+\Delta I]/V$ would be larger and lead to a lower signal to noise ratio.

Now, consider a two-dimensional array of $N$ nanowires as shown in Fig. 3. Such an array can be fabricated using a variety of techniques involving lithography, self-assembly and epitaxial growth [19-24]. The advantage of using a multi-wire sensor is that the signal current in an $N$-wire array is $N$ times that in a single wire, while the noise current is √$N$ times that in a single wire since the conductance is proportional to $N$. Therefore, the signal-to-noise ratio in an $N$-wire array will be √$N$ times higher than in a single wire. Accordingly,

$$(SNR)_N = \frac{(Ne^2/h)^{1/2}\sqrt{\varepsilon_l\varepsilon_r}\sin\left(\frac{\Delta\theta}{2}\right)\sqrt{1+\tanh\left(\frac{E_F}{2kT}\right)}V}{2\sqrt{2kT\Delta f}}. \quad (29)$$

We are now in a position to determine the smallest change in angle $\Delta\theta = \Delta\theta_1 - \Delta\theta_2$ that an $N$-wire sensor can detect. Let us assume that we want the sensor area to be 1 cm² and since one can easily fabricate a parallel array of wires with a density of $10^{10}$/cm², we will have $N = 10^{10}$. We will assume that the semiconductor is InAs. Since the Fermi level in this material is usually pinned within the conduction band, one typically gets a high carrier concentration. Let us conservatively assume that the carrier concentration is $10^6$/cm which yields that $E_F = 30$ mV so that the factor $\sqrt{\left[1+\tanh\left(\frac{E_F}{2kT}\right)\right]}$ is 1.23 at room temperature. We will also assume that the spin injection and detection efficiencies are 70% since that has been demonstrated at room temperature [25]. Furthermore, we want the sensor to be low power, dissipating no more than 1 Watt/cm². The last condition means that $NGV^2 = 1.23N\left(e^2/4h\right)\left[1-\varepsilon_i\varepsilon_2\right]V^2 = 1$ watt, so that $V = 4$ mV. Therefore, in order to maintain a signal-to-noise ratio of at least unity (0 db) in the $N$-wire array, we will need

$$\sin\left(\frac{\Delta\theta_{min}}{2}\right) = \frac{2\sqrt{2kTh\Delta f}}{e\sqrt{N\varepsilon_i\varepsilon_r\left\{1+\tanh\left(\frac{E_F}{2kT}\right)\right\}}V}$$

$$\Delta\theta_{min} \approx \frac{4\sqrt{2kTh\Delta f}}{e\sqrt{N\varepsilon_i\varepsilon_r\left\{1+\tanh\left(\frac{E_F}{2kT}\right)\right\}}V}. \qquad (30)$$

The above condition gives the minimum $\Delta\theta = \Delta\theta_1 - \Delta\theta_2$ that we can detect. According to this condition, $\Delta\theta \geq 1.7\times10^{-10}$ radians/√Hz at room temperature, if we limit the bias voltage $V$ to 4 mV so that we do not dissipate any more than 1 Watt/cm². A larger bias would allow us to detect a smaller $\Delta\theta$ and hence a smaller stress or strain, but at the expense of increased power dissipation. Consequently, the proposed sensor is able to detect stresses and strains that rotate the magnetization by ~$10^{-10}$ radians/√Hz with a power dissipation of 1 Watt/cm$^2$.

We now proceed to calculate the stress and strain which will produce a rotation of $10^{-10}$ radians. Equation (9) shows that $\Delta\theta$ varies linearly with stress, which results in a *linear* sensor with minimal distortion. The variation in $\Delta\theta$ with applied stress is

$$\frac{\partial(\Delta\theta_1)}{\partial\sigma_1} = \frac{(3/2)\lambda_s^{Nickel}}{\mu_0 H_{applied} M_s^{Nickel}}; \quad \frac{\partial(\Delta\theta_2)}{\partial\sigma_2} = \frac{(3/2)\lambda_s^{LSMO}}{\mu_0 H_{applied} M_s^{LSMO}} \quad (31)$$

and the variation with applied strain $\varepsilon$ (assuming both contacts are subjected to the same strain) is obtained from

$$\frac{\partial(\Delta\theta)}{\partial\varepsilon} = \frac{\partial(\Delta\theta_1)}{\partial\sigma_1}\frac{\partial\sigma_1}{\partial\varepsilon} - \frac{\partial(\Delta\theta_2)}{\partial\sigma_2}\frac{\partial\sigma_2}{\partial\varepsilon} = -\left[Y_{Nickel}\frac{(3/2)|\lambda_s^{Nickel}|}{\mu_0 H_{applied} M_s^{Nickel}} + Y_{LSMO}\frac{(3/2)|\lambda_s^{LSMO}|}{\mu_0 H_{applied} M_s^{LSMO}}\right] \quad (32)$$

where $Y$ is the Young's modulus of the material. We can restate Equation (32) as

$$\frac{\partial\varepsilon}{\partial(\Delta\theta)} = -\left[\frac{(3/2)|\lambda_s^{Nickel}|Y_{Nickel}}{\mu_0 H_{applied} M_s^{Nickel}} + \frac{(3/2)|\lambda_s^{LSMO}|Y_{LSMO}}{\mu_0 H_{applied} M_s^{LSMO}}\right]^{-1} \quad (33)$$

which will allow us to find the stress and the strain which will misalign the contacts' magnetizations by an angle $\Delta\theta$.

For polycrystalline nickel contacts, the following values may be used to get an estimate of the stress and strain sensitivity [25]: $(3/2)\lambda_s^{Nickel}$ =-3×10$^{-5}$, $Y_{Nickel}$ = 2×10$^{11}$ Pa, and $M_s^{Nickel}$ = 4.84 × 10$^5$ A/m. These parameters for LSMO are of the same order of magnitude [9], except that the sign of the magnetostriction coefficient is positive [9]. Assuming $H_{applied}$ = 200 Oe = 1.6 ×10$^4$

A/m, we get $\partial\varepsilon/\partial(\Delta\theta)$ = -0.81×10$^{-3}$/radian. Therefore, the room temperature strain sensitivity of this sensor is ~ 10$^{-13}$/√Hz, with a sensing area of ~ 1 cm$^2$ and power dissipation of ~ 1 watt.

It should be noted that the elastic response of nanostructures [26] could be affected by the substrate and pre-stress may result due to interface misfits [26]. However, modeling such effects is beyond the scope of this work. Furthermore, since the nano-pillars are perpendicular to the substrate and also sense stress perpendicular to the substrate plane, these factors will have minimal effect on the sensor performance.

Finally, there is one last consideration, i.e. whether a field of 200 Oe is sufficient to maintain the contacts magnetized in the 45$^0$ orientation at room temperature. The magnetization energy $\mu_o H_{applied} M_s v$ for either contact at this magnetic field is about 21 $kT$ at room temperature, if we assume that $v$ = 30 nm × 10 nm × 30 nm. Hence the magnetic field is sufficiently strong to keep the contacts magnetized collinear with the magnetic field at room temperature. Furthermore, Equation (16) shows that the angle $\alpha = (k_2 - k_1)L = g\mu_B \mu_0 H_{applied} \frac{m^*}{\hbar^2 k_F} L = g\mu_B \mu_0 H_{applied} \frac{\sqrt{m^*}}{\hbar\sqrt{2E_F}} L$. Assuming that $L$ = 1 μm, $m^*$ is 0.03 times the free electron mass (InAs), $E_F$ = 30 meV, and g =15 (InAs), $\alpha$ = 2.5$^0$, which is negligible. This justifies our ignoring the last term in Equation (27).

**IV. Conclusion**

In conclusion, we have proposed a spintronic strain/stress sensor with ultrahigh sensitivity which is low power and miniaturizable. Only two basic effects are required to ensure that the proposed sensor performs as claimed: (1) rotation of magnetization of a magnetostrictive magnet by strain which has been experimentally demonstrated, although not in single domain nanomagnets [27], and (2) the sensitivity of the current in a spin valve to the relative orientations of the contacts. The latter is well-established and forms the basis of tunneling magnetoresistance devices which are the staple of magnetic read heads. Therefore, both basic ingredients of the sensor have been experimentally demonstrated.

If we reduce the sensing area to 100 μm × 100 μm, it will reduce the sensitivity by a factor of 100 since the sensitivity scales with the square-root of the number of nanowires. Such a miniaturized sensor will still have a strain sensitivity of ~ $10^{-11}/\sqrt{Hz}$ at room temperature, which is still very competitive. The power dissipation will reduce to 100 μwatts, allowing it to be integrated on a chip.

Spintronic sensors have not found much direct application in sensing mechanical variables such as stress and strain, but based on the analysis presented here, it seems to be a promising route. The concept proposed here differs from existing work on using GMR/TMR structures for strain sensing [28] since we exploit the one-dimensional nature of the nanowire that is necessary to ensure no spin relaxation occurs. This possibly leads to much higher sensitivities than thin film based strain sensing [28].

In calculating the signal-to-noise ratio we had assumed that the signal current $\Delta I$ is larger than the standby current $I(0)$. That latter is exactly zero if the spin injection and detection efficiencies are 100%. However if the efficiencies are low, then it is possible that $I(0) \gg \Delta I$. In that case, the signal-to-noise ratio calculations are invalidated and the sensitivity will be considerably lower. This shows that spin injection and detection efficiencies are very critical for this sensor. One possible way to circumvent this problem is to have a long nanowire which can be viewed as a series of phase coherent wires. Each one will have a Landauer resistance of $h/e^2$. That will decrease $I(0)$ commensurately, but not affect $\Delta I$ since the stress is same in all of them. This method may ultimately make $\Delta I < I(0)$ and improve the sensitivity

**Figure Captions**

Fig. 1: A quantum wire with ferromagnetic contacts. Current flows along the length of the wire (x-direction) when an electrostatic potential is applied between the two contacts. In the absence of any stress, the contacts are magnetized in the x-z plane bearing an angle of $45^0$ with either the x- or the z-axis. The contacts have a square shape in the x-z plane and the thickness (dimension in the y-direction) is much smaller than the linear dimensions (length and width) in the x-z plane. The white arrows show the orientation of the majority spin in either contact. Application of stress makes the magnetizations of the nickel and LSMO contacts rotate by angles $\Delta\theta_1$ and $\Delta\theta_2$ which have opposite signs.

Fig. 2: Energy dispersion relation in the wire in the presence of the applied magnetic field.

Fig. 3: (Top) A compressive stress is applied to an array of free-standing quantum wires along their length. (Bottom) The strain measurement configuration where the sensor is integrated on the test surface.

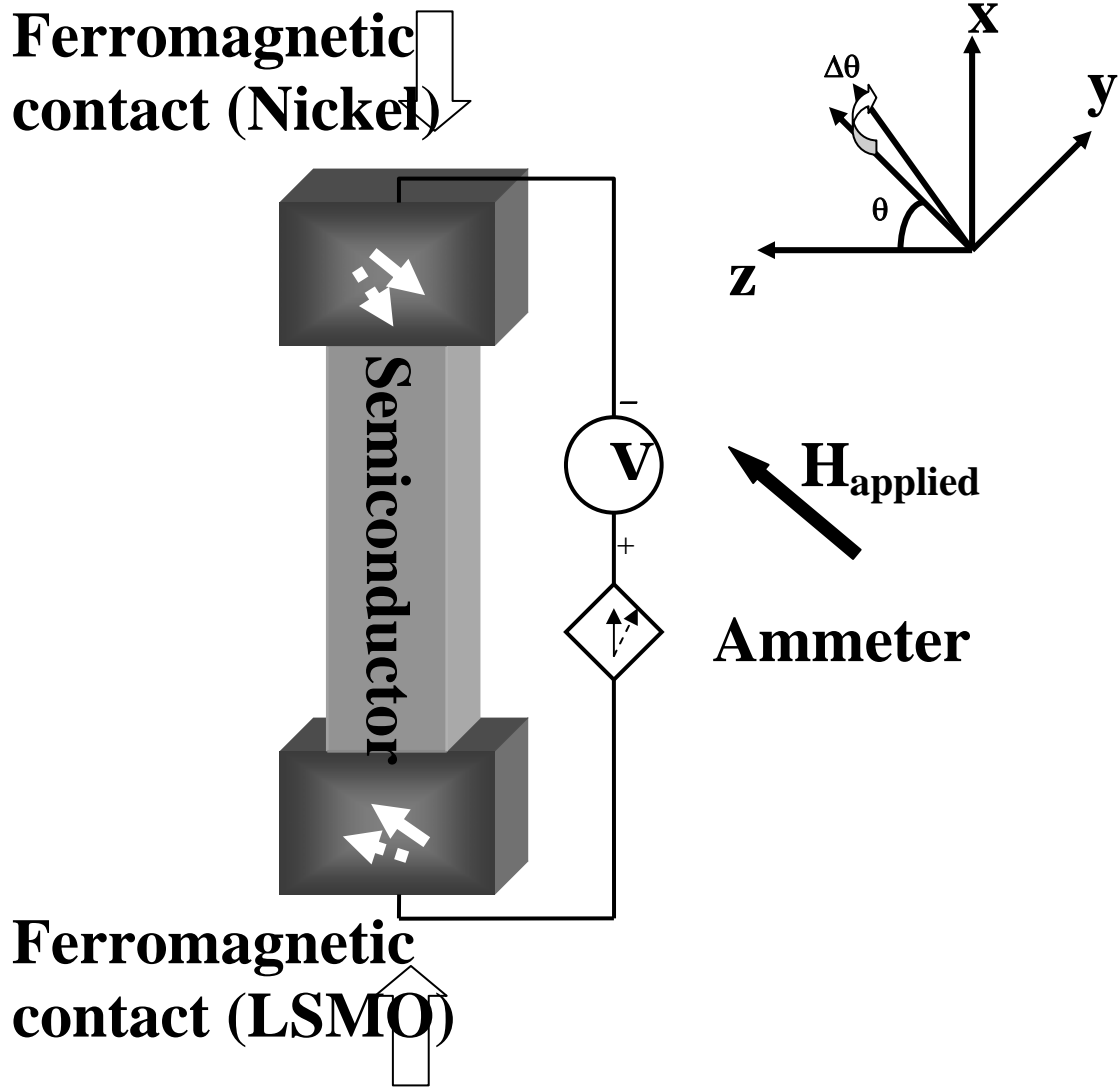

Fig. 1

Atulasimha and Bandyopadhyay

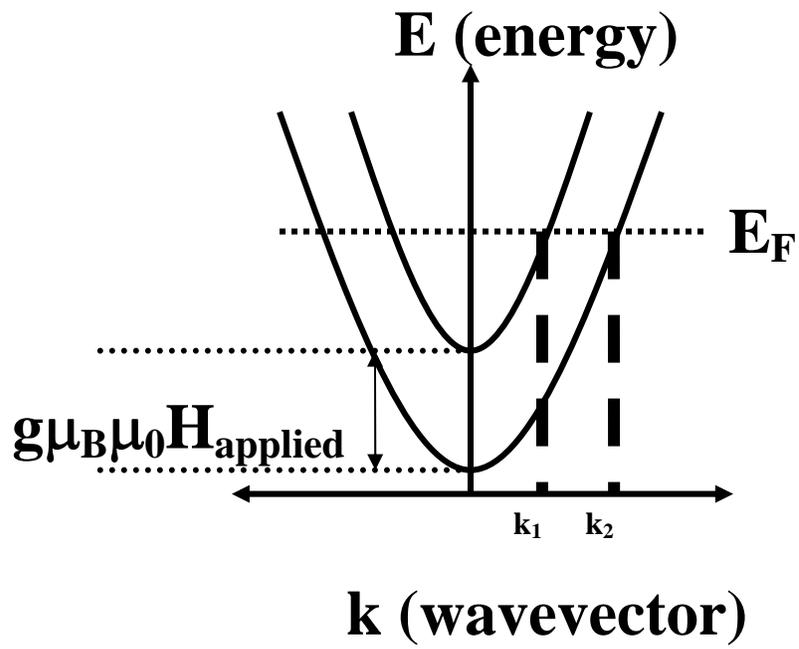

Fig. 2

Atulasimha and Bandyopadhyay

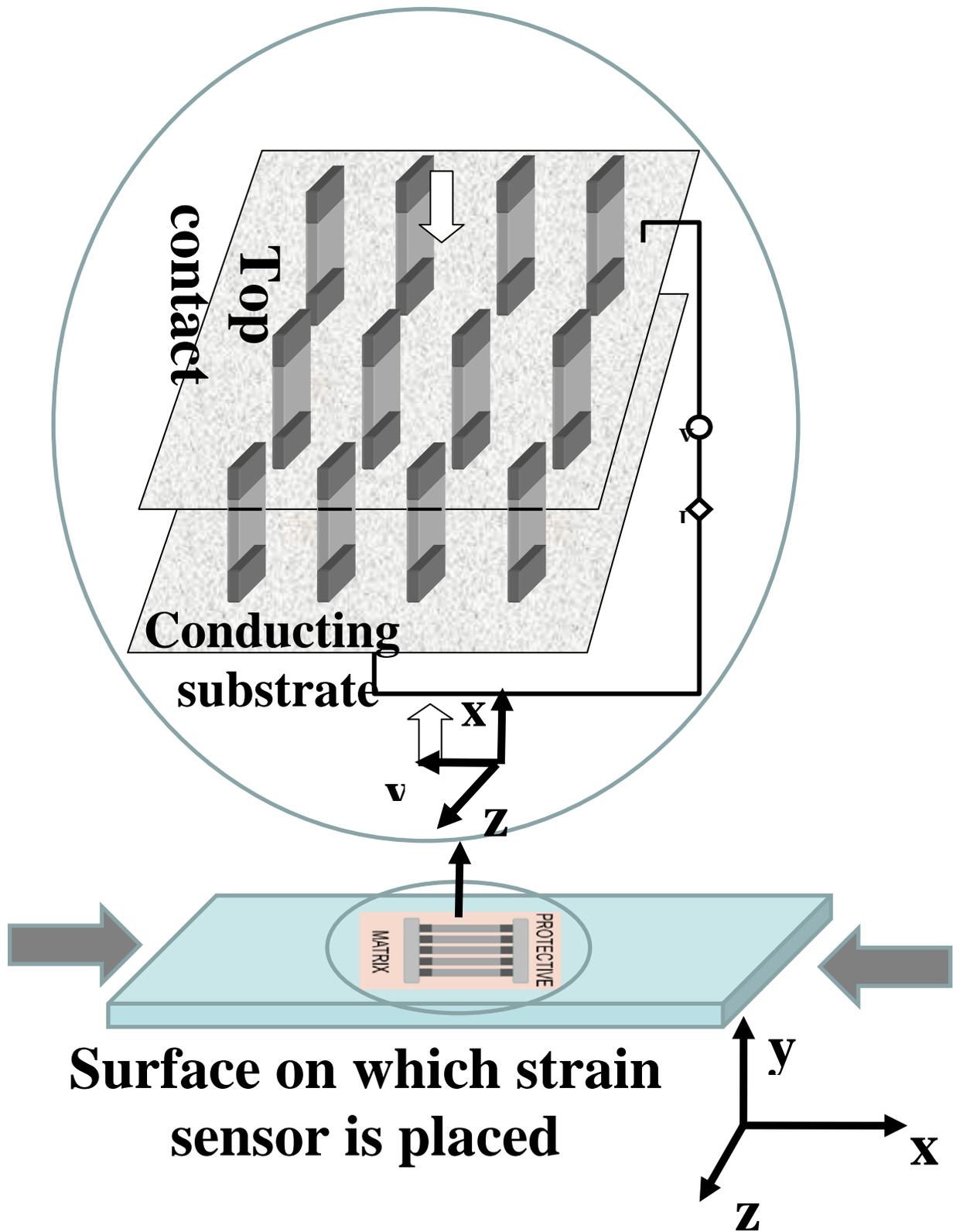

Fig. 3: Atulasimha and Bandyopadhyay